\documentclass[sigconf]{acmart}
\usepackage{graphicx}
\usepackage{amsmath}
\usepackage{booktabs}
\usepackage{algorithm}
\usepackage{algorithmic}
\usepackage{amsfonts}
\usepackage{multirow}
\usepackage{makecell}
\usepackage{subfigure}
\usepackage{color}
\usepackage{caption}
\usepackage{bm}
\usepackage{epstopdf}
\usepackage{url}
\usepackage[cal=cm]{mathalfa}
\usepackage{balance}
\usepackage{threeparttable}
\usepackage{lipsum}
\usepackage{enumitem}
 \usepackage{amsmath}

\usepackage{xcolor}

\AtBeginDocument{%
  \providecommand\BibTeX{{%
    \normalfont B\kern-0.5em{\scshape i\kern-0.25em b}\kern-0.8em\TeX}}}

\copyrightyear{2025} 
\acmYear{2025} 
\setcopyright{cc}
\setcctype{by}
\acmConference[CIKM '25]{Proceedings of the 34th ACM International
Conference on Information and Knowledge Management}{November 10--14,
2025}{Seoul, Republic of Korea}
\acmBooktitle{Proceedings of the 34th ACM International Conference on
Information and Knowledge Management (CIKM '25), November 10--14, 2025,
Seoul, Republic of Korea}\acmDOI{10.1145/3746252.3761015}
\acmISBN{979-8-4007-2040-6/2025/11}

\begin{document}
\title{MARM: Unlocking the Recommendation Cache Scaling-Law through Memory Augmentation and Scalable Complexity}
\renewcommand{\shorttitle}{MARM}

\author{Xiao Lv}
\affiliation{
  \institution{Kuaishou Technology}
  \city{Beijing}
  \country{China}
}
\email{lvxiao03@kuaishou.com}

\author{Jiangxia Cao}
\affiliation{
  \institution{Kuaishou Technology}
  \city{Beijing}
  \country{China}
}
\email{caojiangxia@kuaishou.com}

\author{Shijie Guan}
\affiliation{
  \institution{Kuaishou Technology}
  \city{Beijing}
  \country{China}
}
\email{guanshijie@kuaishou.com}
 
\author{Xiaoyou Zhou}
\affiliation{
  \institution{Kuaishou Technology}
  \city{Beijing}
  \country{China}
 }
\email{zhouxiaoyou@kuaishou.com}

\author{Zhiguang Qi}
\affiliation{
  \institution{Kuaishou Technology}
  \city{Beijing}
  \country{China}
}
\email{qizhiguang@kuaishou.com}

\author{Yaqiang Zang}
\affiliation{
  \institution{Kuaishou Technology}
  \city{Beijing}
  \country{China}
 }
\email{zangyaqiang@kuaishou.com}

\author{Ming Li}
\affiliation{
  \institution{Kuaishou Technology}
  \city{Beijing}
  \country{China}
}
\email{liming03@kuaishou.com}

\author{Ben Wang}
\affiliation{
  \institution{Kuaishou Technology}
  \city{Beijing}
  \country{China}
}
\email{wangben@kuaishou.com}

\author{Kun Gai}
\affiliation{
  \institution{Independence}
  \city{Beijing}
  \country{China}
}
\email{gai.kun@qq.com}

 \author{Guorui Zhou}
 \authornote{Corresponding authors}
\affiliation{
  \institution{Kuaishou Technology}
  \city{Beijing}
  \country{China}
}
\email{zhouguorui@kuaishou.com}

\begin{abstract}
  Scaling-law has guided the language model design for past years, e.g., GPTs, enabling the estimation of expected model performance with respect to the size of learnable parameters and the scale of training samples.
It is worth noting that the scaling laws of NLP cannot be directly applied to recommendation systems due to the following reasons: 
(1) The amount of training samples and model parameters is typically not the bottleneck for the model. Our recommendation system can generate over 50 billion user samples daily, and such a massive amount of training data can easily allow our model parameters to exceed 200 billion, surpassing many LLMs (about 100B). 
(2) It is essential to control FLOPs carefully in recommendation system. In training, we need to process a vast number of recommendation samples every day. During online inference, we must respond within milliseconds (LLMs usually take a few seconds).
Considering the above differences with LLM, we can conclude that: for a RecSys model, compared to model parameters, the FLOPs is a more expensive factor that requires careful control.

In this paper, we propose our milestone work, \textbf{MARM} (Memory Augmented Recommendation Model), which explores a new cache scaling-law successfully.
By caching part of complex module calculation results, our MARM extends the single-layer attention-based sequences interests modeling module to a multi-layer setting with minor inference FLOPs cost (i.e, module time complexity $\mathcal{O}(n^2*d) \rightarrow \mathcal{O}(n*d)$).
Equipped with the cache idea, our MARM solution significantly overcomes computational bottlenecks and can seamlessly empower all interest extraction modules for user sequences, and even other models. 
To support our MARM, we construct a 60TB cache storage center for offline training and online serving.
Comprehensive experiment results show that our MARM brings offline 0.43\% GAUC improvements and online 2.079\% play-time per user gains.
Our MARM has been deployed on a real-world short-video platform, serving tens of millions of users daily.

\end{abstract}


\maketitle

\section{Introduction}
\begin{figure}[t!]
  \centering
  \includegraphics[width=8cm,height=11cm]{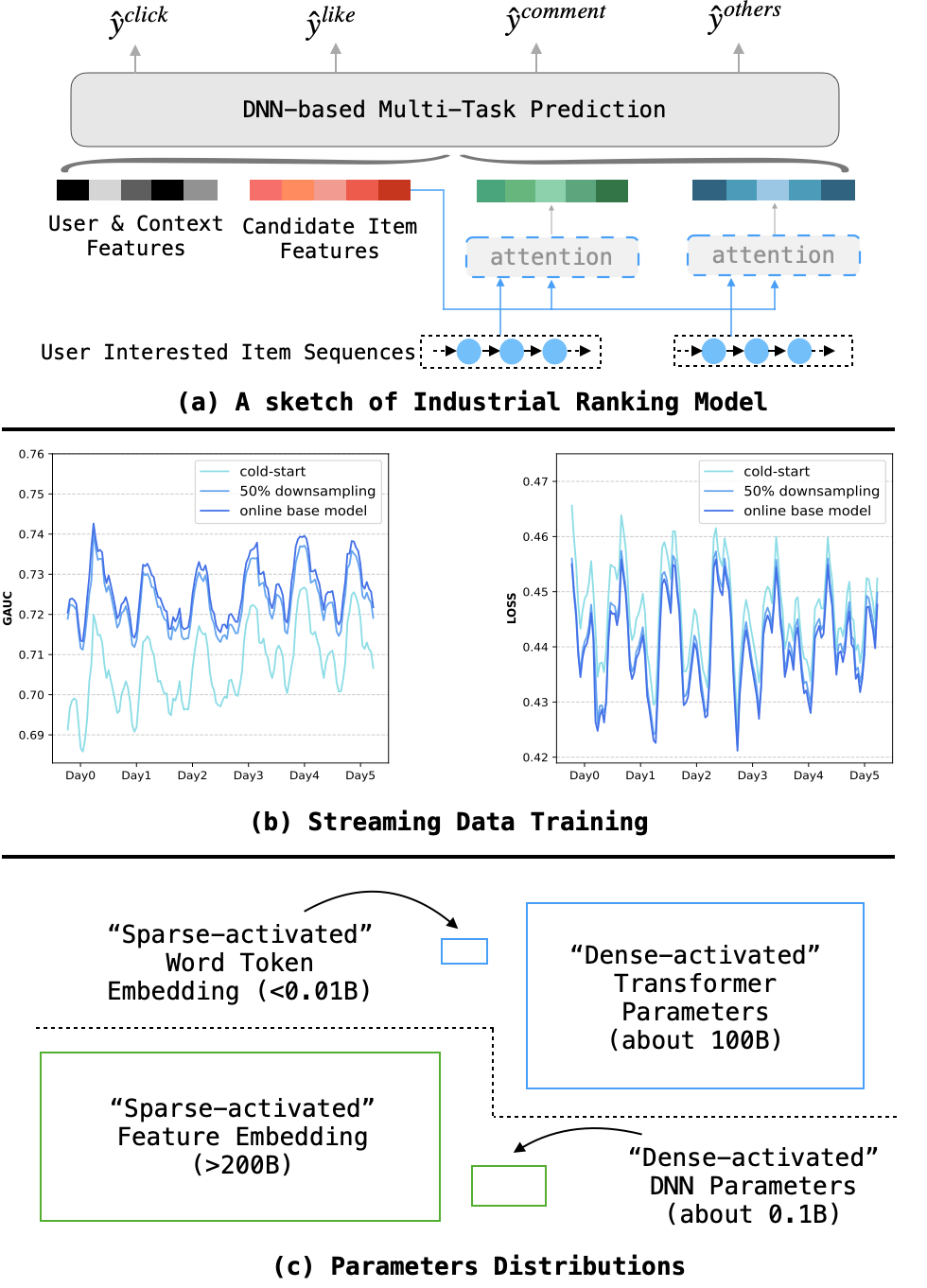}
  \caption{(a) A toy example of ranking model. (b) Performance gap in Streaming Data Training. Warming up from the online base model, a 50\% sample downsampling leads to a continuous decline in model performance. 
  (c) Parameter distributions. The parameters of an LLM model (blue) are mainly located in `dense-activated' transformer-based DNN module, while ranking models (green) have a parameter distribution primarily composed of `sparse-activated' features.}
  \label{fig:intro}
\end{figure}

\begin{figure*}[t]
  \centering
  \includegraphics[width=18cm,height=10cm]{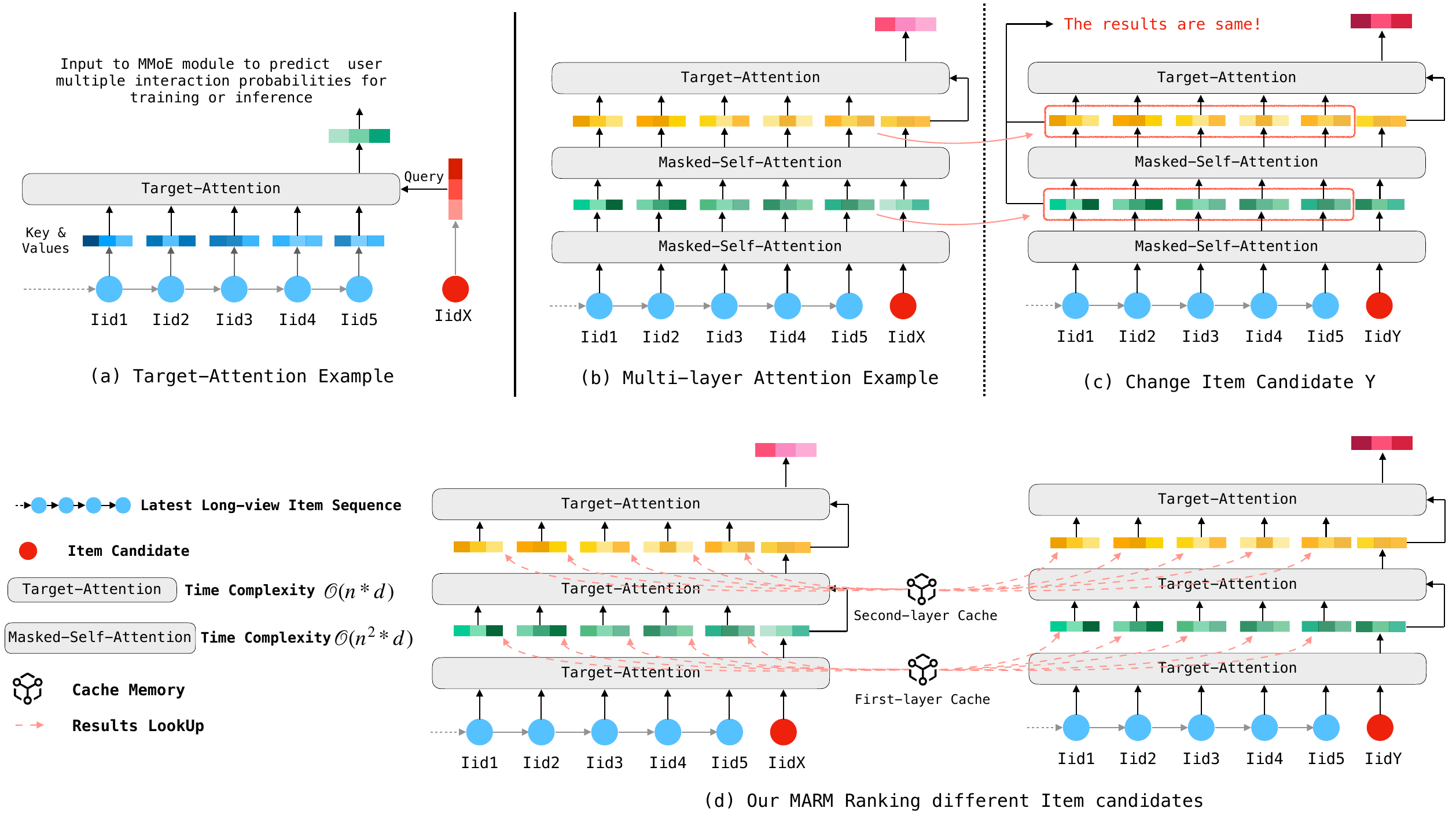}
  \caption{Motivation of MARM, using caching idea to reduce higher-FLOPs self-attention to lower-FLOPs target-attention.}
  \label{fig:targetattention}
\end{figure*}

%
In recent years, the scaling of model size and data-scale has been a vital point in various fields, e.g., NLP, CV, and RecSys.
%
According to OpenAI Scaling-laws technique report\cite{kaplan2020scaling}, as the volume of data and the breadth and depth of models increase, the performance of the models follows a certain power-law improvement. 
Following the excited Scaling-laws tenet, many Transformer-based large models are proposed and achieve remarkable performance, e.g., ChatGPT for conversation, Qwen-VL\cite{bai2023qwen} for multi-modal understanding, DeepSeek-Coder\cite{guo2024deepseek} for code generation, and so on.
There are also some efforts to validate Scaling-laws in RecSys area (e.g., Wukong\cite{zhang2024wukong}, HSTU\cite{zhai2024actions}), but these methods make strong assumptions on feature engineering or model architectures, such as removing all static features or replacing entire model modules with Transformers.
Actually, the common wisdom always formed RecSys ranking model designing as follows: first crafting hundreds of elaborated features as model input and then using a multi-task prediction module to obtain the click/like/comment/others scores for a user and item candidate pair (in Figure~\ref{fig:intro}(a)).
Nevertheless, recent RecSys Scaling-laws studies are greatly changing such learning paradigms, making it difficult to deploy them in real online RecSys ranking models directly.
In this paper, we focus on exploring the Scaling-laws under the above widely-used ranking model architecture.
Due to differences in ranking model architecture and usage with pure Transformer-based LLMs, we conclude three reasons why NLP Scaling-laws do not match to RecSys ranking model:
\setlist{nosep}
\begin{itemize}[leftmargin=*,align=left]
    \item \textbf{For the training samples}: different from NLP models having a stable data corpus to train LLM from scratch, the industrial RecSys model always follows a streaming training paradigm.
    %
    At real-world RecSys, our model is streaming trained over 50 billion user logs daily. Here we give three model variants training details: the `online base model', `online base model with 50\% training data downsampling', and `cold-start', where the `online base model' is trained over years.
    From Figure~\ref{fig:intro}(b), we can draw conclusions that:
    (1) Training from scratch will hurt performance significantly; (2) although the model parameters are warm-started from a several-year well-trained model, down-sampling the real-time training samples will still have performance degradation.
    Such phenomena validate that the RecSys model is data-hungry for \textbf{infinite} data to capture real-time users' preferences. Therefore, the analysis of training data-scale is unnecessary in industry.
    %
   %
   %
    %
    \item \textbf{For the learning parameters}: As shown in Figure~\ref{fig:intro}(c), LLM parameters are mainly located in `dense-activated' Transformer-based DNN module (about 100 Billion), while the `sparse-activated' word token parameters are small (<0.01B). 
    Ranking models always have opposite parameter distribution: major learnable parameters concentrate on `sparse-activated' feature parameters (\textbf{>200B}) while `dense-activated' DNN parameters (about 0.1B) are much smaller.
    %
    Where `sparse' means only small part parameters will activate to involve calculating (i.e., we only need to look up few words tokens for a sentence in LLM; we only need to look up one user/item ID for a user-item training sample in RecSys), and `dense' means the calculation flow parameters which will fully activate to estimate final results.
    Indeed, in terms of the amount of learnable parameters, our ranking model (>200B) far exceeds many LLMs (around 100B), which indicates that learnable parameters are not the bottleneck of ranking models.
    \item \textbf{For the computational complexity, FLOPs}: Actually, RecSys model inference FLOPs is much lower than LLM and has upper bound.
    This is because our model needs to process hundreds of millions of requests efficiently, and that's why our dense-activated DNN module is much smaller (about 0.1B).
    Thus, to ensure our service's stability and robustness when processing each request in \textbf{milliseconds} (LLMs usually take seconds), we cannot blindly increase inference complexity. Additionally, since the real-time nature of training recommendation models greatly affects the quality of online performance, the computational complexity of model under limited offline training resources will impact the real-time nature of streaming training, resulting in trade-off with online performance, e.g, if the offline model has 20 minutes delay, it will cause 1\% drop in online performance.
\end{itemize}

Based on the analysis of the above three aspects, compared to LLM, we can observe that our ranking model has its \textbf{advantages} and \textit{disadvantages}: (1) \textbf{`unlimited' streaming data}, (2) \textbf{massive parameter storage}, (3) \textit{relatively low-FLOPs DNN module}.
In other words, for ranking model, data and storage resources are relatively cheap, but computing resources are cautiously expensive.
Motivated by these points, we consider whether we can use the model's strengths in data and storage to compensate for its weakness in computing.
In other words, \textbf{can we cache part of complex module calculation results to degenerate its time complexity?}

To answer the question, we propose our milestone work, MARM (Memory Augmented Recommendation Model), which achieves a new RecSys Scaling-law between the cache size and model performance.
In MARM, we extend one of the most important user interests extraction modules in industry ranking model, which aims to calculate users' historical item importance with candidate items, as shown at the bottom of Figure~\ref{fig:intro}(a).
%
To the best of our knowledge, in implementing this module, many outstanding methods (e.g., DIN\cite{zhou2018deep}, SIM\cite{pi2020search}, SDIM\cite{cao2022sampling}, TWIN\cite{chang2023twin}) utilize a single-layer \textbf{target-attention (TA)} mechanism, as shown in Figure~\ref{fig:targetattention}(a).
Intuitively, such module could be enhanced by stacking multi-layer \textbf{self-attention (SA)} before the final TA layer, as shown in Figure~\ref{fig:targetattention}(b).
Unfortunately, the time complexity of SA ($\mathcal{O}(n^2*d)$) is much higher than TA ($\mathcal{O}(n*d)$, where $n$ denotes sequence length, $d$ denotes representation dimension), which is more sensitive on longer sequences (e.g., $n > 1000$) and will result in a large increase in FLOPs.
In online-serving, our ranking model needs to simultaneously predict 50$\sim$8000 user-item candidate pairs to find the best dozens of items for a user. This leads to a lot of repeated calculations in the naive multi-layer attention mechanism, which will bring heavy pressure for our system (in Figure~\ref{fig:targetattention}(c)).
%

Fortunately, owing to the temporal order of streaming data widely used in industry RecSys, if we cache the frequently used masked SA module results on the latest samples, the complex masked SA layer can be replaced by simple TA layer to efficiently predict different item candidates well, as shown in Figure~\ref{fig:targetattention}(d).
In this way, our MARM extends the single-layer attention-based sequences interests modeling module to a multi-layer setting with a minor FLOPs cost.
Specifically, our MARM solution significantly overcomes computational bottlenecks and could seamlessly empower all interest extraction modules for user sequences.
Moreover, based on MARM cache results, we find it beneficial for other models, such as retrieval and cascading models.
 In our experiments, we fully explore scaling laws between cache size and model performance, seeking a perfect balance among cached attention depth, sequence length, and embedding dimension. Our main contributions are:
\begin{itemize}[leftmargin=*,align=left]
\item \textbf{We are the first work to explore new scaling-laws under the widely-used ranking model architecture from a fresh cache-performance perspective}.
\item We devise a simple-yet-efficient approach, MARM, which utilizes the strengths of ranking models in data and storage to reduce high time complexity of masked-self-attention. We successfully extend the single-layer attention-based sequences interests modeling module to a multi-layer setting with minor inference FLOPs. 
\item We show our MARM is highly adaptable and scalable, seamlessly integrating into existing high-performance transformer-based models, and facilitating a smooth transition towards GPT-style models. Besides, extensive experiments and comparative ablation studies were conducted offline\&online (average improves 0.43\% GAUC offline \& 2.079\% play-time per user online). 
\end{itemize}


\begin{figure}[t]
  \centering
  \includegraphics[width=8cm,height=16cm]{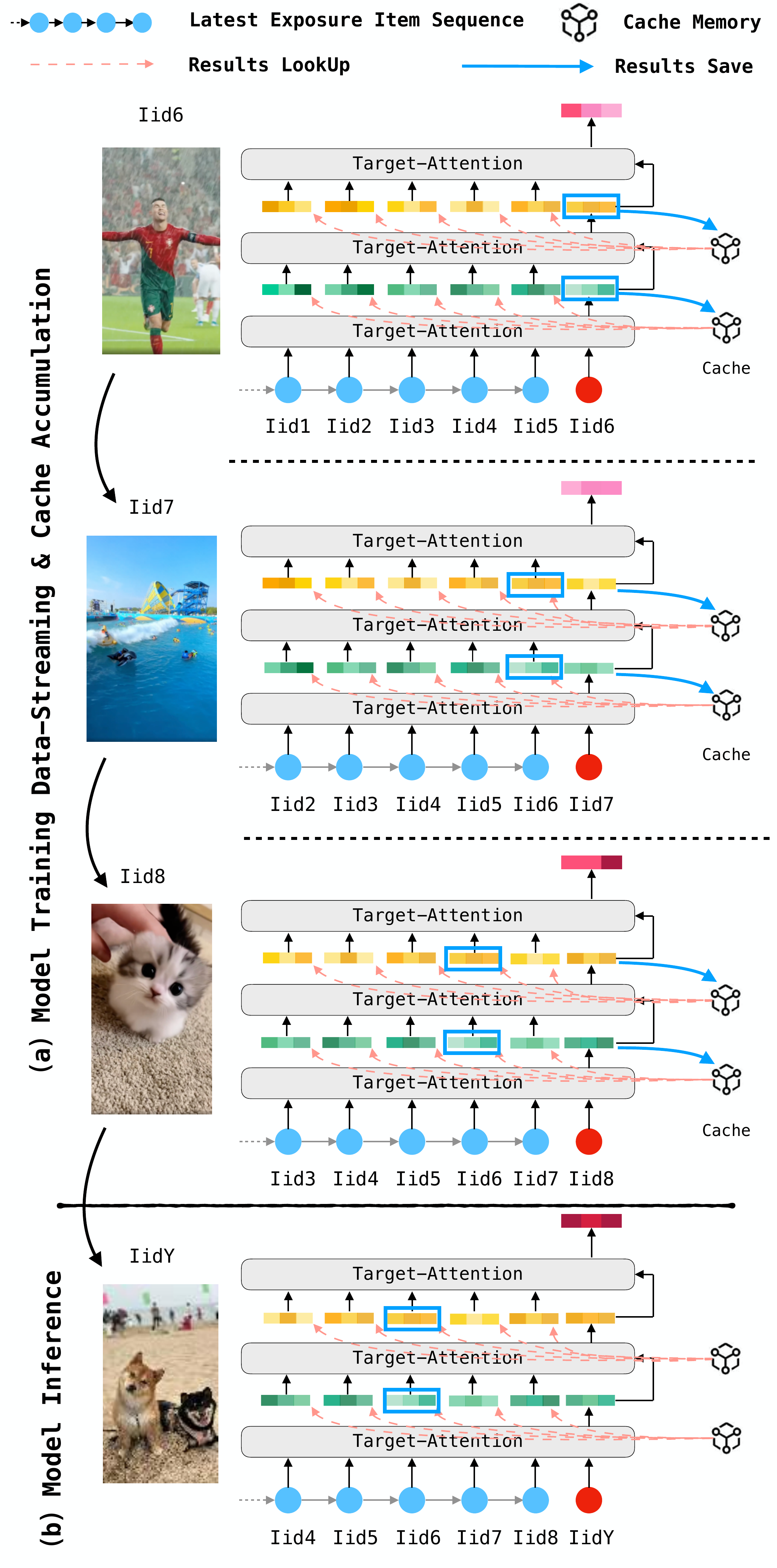}
  \caption{MARM realtime cache update\&use workflow.}
  \label{fig:cacheaccumulation}
\end{figure}

\section{Methodology}

In this section, \textbf{we do not explain the total details of our ranking model's architectures or losses}.
Specifically, for better understanding, we only dive into the MARM module and explain how it works, as sketched in Figure~\ref{fig:targetattention}(d).
%

%
%

\subsection{MARM Workflow}\label{sec:workflow}
\label{workflow}

This section describes MARM workflow under the latest exposure item sequences, which builds a multi-layer decoder-only Transformer architecture to capture user interests with respect to a certain target item.
Our MARM mainly consists of four parts:
\begin{enumerate}[leftmargin=*,align=left]
    \item A \textbf{sequence generator} to produce items exposed to user.
    \item An external cache memory storage to \textbf{look up} results.
    \item \textbf{Multi-layer target-attention} module to calculate results.
    \item \textbf{Sending} the intermediate results to the cache store.
\end{enumerate}


\subsubsection{Sequence Generator}
Given arbitrary user ID $\texttt{UidX}$, we \textbf{assume} that our sequence generator could produce his/her latest exposure item sequence in chronological order as:
%
%
\begin{equation}
\small
\begin{split}
[\texttt{Iid1}, \dots, \texttt{Iidn}] = \texttt{ExposureSeqGen}(\texttt{UidX}, n).
\end{split}
\label{seqgen}
\end{equation}
%
%
where $n$ denotes sequence length, the $[\texttt{Iid1}, \dots, \texttt{Iidn}]$ is the behavior sequence for user $\texttt{UidX}$.
Of course, since our sequence has clear user feedback, we can also choose to use items that meet certain conditions. For example, we might only use items that have a long-view label, see Section~\ref{sec:sim} for this extension.

\subsubsection{Cache Memory Storage Look-Up}
As shown in Figure~\ref{fig:targetattention}, given \textbf{\texttt{UidX}'s sequence} $[\texttt{Iid1}, \dots, \texttt{Iidn}]$ and \textbf{item candidate} \texttt{IidY}, there are two types of representations needed to support our MARM: the learnable ID-based embedding and un-learnable cached results.
%
\begin{enumerate}[leftmargin=*,align=left]

    \item \textbf{Learnable embedding}: 
    In RecSys, ID-based embedding is a part of model sparse-activated feature that can look up directly:
\begin{equation}
\small
\begin{split}
\mathbf{IidY} = &\texttt{SparseFeatureLookUp}(\texttt{IidY}),\\
[\mathbf{Iid1}, \dots,\mathbf{Iidn}] =& \texttt{SparseFeatureLookUp}([\texttt{Iid1}, \dots, \texttt{Iidn}]).
\end{split}
\label{sparse_embedding}
\end{equation}
where $\mathbf{IidY}\in \mathbb{R}^F$, $F$ is a hyper-parameter to control the feature dimension.
The $[\mathbf{Iid1}, \dots, \mathbf{Iidn}]$ is the the bottom target-attention input in Figure~\ref{fig:targetattention}(a).
%
%
It is worth noting that viewed items could fulfill some other attributes, such as author ID, tags, and user interaction labels.
In this paper, for simplicity, we will only retain the item ID to represent them.
\item \textbf{Un-Learnable Cached results}: Considering we have cached $L$-layer intermediate results in an external key-value memory storage, we need to look up those results precisely.
%
%
Therefore, we devise a hash strategy to generate `key' to reflect the meaning: cached value of user-item pair at layer $i$.
For example, given user \texttt{UidX}, and its item sequence [\texttt{Iid1},\dots], to find cache results at layer depth $i$, we could generate hash key list as:
\begin{equation}
\small
\begin{split}
[\texttt{Iid1}^i_{\texttt{UidX}},\dots] = \texttt{UserItemDepthHash}(\texttt{UidX}, [\texttt{Iid1},\dots], i).
\end{split}
\label{hash_key}
\end{equation}
where the $\texttt{Iid1}^i_{\texttt{UidX}}\in \mathbb{Z}$ denotes the Hash-key to access the correct unique cache result.
Based on them, we have:
\begin{equation}
\small
\begin{split}
[\mathbf{Iid1}^i_{\texttt{UidX}},\dots] = \texttt{MARMCacheLookUp}([\texttt{Iid1}^i_{\texttt{UidX}},\dots])
\end{split}
\label{cachelookup}
\end{equation}
where $\mathbf{Iid1}^i_{\texttt{UidX}} \in \mathbb{R}^d$, and $d$ is attention module dimension.
As a result, combining all layers' cache results, we have:
%
$$
\begin{footnotesize}
\begin{bmatrix}
\mathbf{Iid1}^1_{\texttt{UidX}}& \mathbf{Iid2}^1_{\texttt{UidX}}& \mathbf{Iid3}^1_{\texttt{UidX}}& \dots& \mathbf{Iidn}^1_{\texttt{UidX}} \\
\mathbf{Iid1}^2_{\texttt{UidX}}& \mathbf{Iid2}^2_{\texttt{UidX}}& \mathbf{Iid3}^2_{\texttt{UidX}}& \dots& \mathbf{Iidn}^2_{\texttt{UidX}} \\
\vdots& \vdots& \vdots& \vdots& \vdots \\
\mathbf{Iid1}^L_{\texttt{UidX}}& \mathbf{Iid2}^L_{\texttt{UidX}}& \mathbf{Iid3}^L_{\texttt{UidX}}& \dots& \mathbf{Iidn}^L_{\texttt{UidX}}
\label{cache_results}
\end{bmatrix}
\end{footnotesize}
$$

\end{enumerate}

\begin{figure*}[t]
  \centering
  \includegraphics[width=18cm,height=4.2cm]{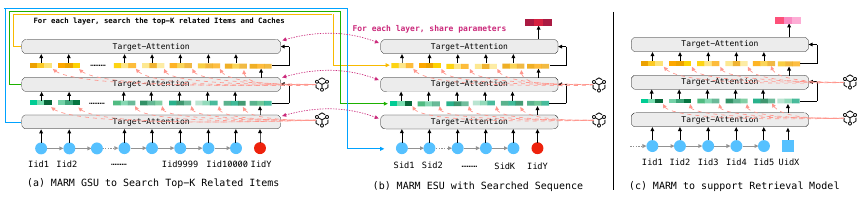}
  \caption{Using the MARM framework to handle long sequences with SIM GSU/ESU; Example to support Retrieval Model.}
  \label{fig:marm_extension}
\end{figure*}

\subsubsection{Multi-Layer Target-Attention}
Up to now, we have obtained learnable target-attention (TA) ID embeddings and subsequent unlearnable TA cached results.
%
%
According to learnable embeddings, we first calculate them with a certain target item information $\mathbf{IidY}$ through TA mechanism:
\begin{equation}
\small
\begin{split}
\mathbf{IidY}_{\texttt{UidX}}^1& = \texttt{Target-Attention}(\mathbf{IidY}, [\mathbf{Iid1}, \dots]),\\
\end{split}
\label{marm_attention_learn}
\end{equation}
Next, we feed the results $\mathbf{IidY}_{\texttt{UidX}}^1 \in \mathbb{R}^d$ to calculate with un-learnable cached results to simulate the high-FLOPs masked-self-attention (MSA) ($\mathcal{O}(n^2*d)$) by the low-FLOPs TA ($\mathcal{O}(n*d)$):
\begin{equation}
\small
\begin{split}
\mathbf{IidY}_{\texttt{UidX}}^2 = &\texttt{Target-Attention}^1(\mathbf{IidY}_{\texttt{UidX}}^1, [ \mathbf{Iid1}_{\texttt{UidX}}^1, \dots]),\\
\mathbf{IidY}_{\texttt{UidX}}^3 = &\texttt{Target-Attention}^2(\mathbf{IidY}_{\texttt{UidX}}^2, [\mathbf{Iid1}_{\texttt{UidX}}^2, \dots]),\\
&\dots\dots\dots\dots\\
\mathbf{IidY}_{\texttt{UidX}}^{L+1} = &\texttt{Target-Attention}^{L}(\mathbf{IidY}_{\texttt{UidX}}^{L}, [\mathbf{Iid1}_{\texttt{UidX}}^{L}, \dots]),\\
\end{split}
\label{marm_attention}
\end{equation}
where $\mathbf{IidY}_{\texttt{UidX}}^{L+1}\in \mathbb{R}^d$ denotes final sequences interests modeling output belonging to (\texttt{UidX}, \texttt{IidY}) pair, and will be fed into following multi-task learning module to predict ground-truth labels.
%
%

\subsubsection{Sending results to Cache Memory}
The above sections have introduced our MARM calculating process. Next, we will explain how to save such intermediate results to the cache store.
Similarly, we first generate the `hash key' to indicate the result meanings, e.g., the $(\texttt{UidX}, \texttt{IidY})$ result at different layers $[1, \dots, L]$:
\begin{equation}
\footnotesize
\begin{split}
[\texttt{IidY}_{\texttt{UidX}}^1, \dots, \texttt{IidY}_{\texttt{UidX}}^{L}] = \texttt{UserItemDepthHash}(\texttt{UidX}, \texttt{IidY}, [1, \dots, L]).
\end{split}
\label{hash_key}
\end{equation}
Afterwards, we can send the values with their keys in order:
\begin{equation}
\small
\begin{split}
\texttt{MARMCacheSave}([\texttt{IidY}^1_{\texttt{UidX}},\dots,\texttt{IidY}^{L}_{\texttt{UidX}}], [\mathbf{IidY}_{\texttt{UidX}}^1, \dots, \mathbf{IidY}_{\texttt{UidX}}^{L}])
\end{split}
\label{cacheSave}
\end{equation}
In streaming training, as we continuously save real-time computation results, MARM cache gradually accumulates computation results of each layer of MARM for all users' historical behaviors.
Through natural accumulation of cache, MARM can apply decoder-only Transformer architecture with minor inference FLOPs. 
The pseudo-code of our MARM is shown in Algorithm~\ref{algo:marm}, and Figure~\ref{fig:cacheaccumulation} also provides a user-watching example in streaming setting.

\subsection{MARM with SIM}\label{sec:sim}
In Section~\ref{workflow}, we have discussed a naive MARM setting, modeling the users' interests under the latest watched items sequence.
However, since the sequence length $n$ in Eq.(\ref{seqgen}) is limited (about 100), the earlier stored results will no longer be used once out of range, resulting in a waste of storage resources.
This section further extends MARM with \textbf{Search-based interests model} (SIM)\cite{pi2020search}, to maximize the effectiveness of cached results.
%
%

%
Formally, SIM proposes a two-stage cascading modeling paradigm: (1) introduce a coarsen General Search Unit (GSU) to retrospect user life-long history (e.g. >10000) for target item, and then search the top-K most related item sequences; (2) employ fine-grained Exact Search Unit (ESU) to compress the searched sequence information to obtain precise user interests with respect to target item.
In past years, SIM has been the vital engine for ranking model iteration, and the recent effort is TWIN~\cite{chang2023twin}, which utilizes the shared GSU and ESU modules for two-stage unbiased modeling.
%
%
Inspired by the TWIN, we also consider applying a synchronous GSU module to \textbf{search the top-K related cache results for each layer} (as shown in Figure~\ref{fig:marm_extension}(a)).
%
%
%
%
Given the latest long-term sequence $[\texttt{Iid1}, \dots, \texttt{Iid}10000]$ of user $\texttt{UidX}$, the MARM GSU aims to generate the multi-layer searched input sequences, for example:
\begin{equation}
\footnotesize
\begin{split}
[\texttt{S\_Iid1}^i, \dots, \texttt{S\_IidK}^i]& = \texttt{MARMGSU}(\texttt{IidY}, [\texttt{Iid1}, \dots, \texttt{Iid}10000], K, i), \\
\end{split}
\label{simsequence}
\end{equation}
where the $\texttt{MARMGSU}$ shared same \texttt{Target-Attention} parameters with our ranking model, and $[\texttt{S\_Iid1}^i, \dots, \texttt{S\_IidK}^i]$ are top-K \textbf{searched} items with highest attention weights for candidate \texttt{IidY} at layer $i$.
Combining all layers’ searched keys, we have:
$$
\begin{footnotesize}
\begin{bmatrix}
\texttt{S\_Iid1}^1& \texttt{S\_Iid2}^1& \texttt{S\_Iid3}^1& \dots& \texttt{S\_IidK}^1 \\
\texttt{S\_Iid1}^2& \texttt{S\_Iid2}^2& \texttt{S\_Iid3}^2& \dots& \texttt{S\_IidK}^2 \\
\vdots& \vdots& \vdots& \vdots& \vdots \\
\texttt{S\_Iid1}^{L}& \texttt{S\_Iid2}^L& \texttt{S\_Iid3}^L& \dots& \texttt{S\_IidK}^L
\label{sim_cache_results}
\end{bmatrix}
\end{footnotesize}
$$
where the searched sequence of different layer maybe different, e.g., $\texttt{S\_Iid}1^i \ne \texttt{S\_Iid}1^{i+1}$.
Afterwards, we can look up their corresponding cached results to support our MARM module in Figure~\ref{fig:marm_extension}(b).
%
%

\subsection{MARM to Support Other Models}\label{sec:othermodels}
Actually, the industrial RecSys always introduces several models to filter candidate items layer by layer:
\begin{itemize}[leftmargin=*,align=left]
    \item Retrieval model: without item candidates, only utilizing users' latest interaction items to fetch \textbf{thousands of} items from the full billion-item pool.
    \item Cascading model: a small ranking model, utilizing extremely efficient modules to select top \textbf{hundreds of} item candidates from the thousands generated by the retrieval model.
    \item Ranking model: in a broad sense, it is the most complex model, which finds the best \textbf{dozens} items from hundreds of candidates.
\end{itemize}
Generally, these models are trained using different strategies and are isolated from each other.
\textbf{Therefore, here is an interesting assumption: can our ranking model MARM cache storage contribute to other Retrieval/Cascading models?} 
%
Unsurprisingly, we found our MARM results could support them seamlessly:
\begin{itemize}[leftmargin=*,align=left]
    \item For Retrieval model, since there is no candidate set of items, we use users' latest 200 watched items to look up the cache results from MARM, and then apply the user ID feature to replace item candidates to conduct target-attention as shown in Figure~\ref{fig:marm_extension}(c). 
    \item For Cascading model, we implement the same multi-layer target-attention process as our Ranking model in Eq.(\ref{marm_attention}). 
    It is important to note that in the cascading model, we do not perform re-accumulation of the MARM cache; instead, we directly use the already accumulated MARM cache and only use latest short-term sequences. This approach saves resources while also enhancing the consistency between cascading and ranking models. 
    %
\end{itemize}

\begin{algorithm}[t]
\footnotesize
\caption{MARM Training and cache accumulation procedure.}
\label{algo:marm}
\begin{algorithmic}[1]
\REQUIRE $n$ denotes sequence length, $L$ denotes depth of MARM, $d$ denotes dimension
\ENSURE Updated MARM Cache Storage.
\WHILE {(\texttt{UidX}, \texttt{IidY}, )  from data-streaming} 
\STATE \#\#\#\#\#\#\textbf{SEQUENCE GENERATION}\#\#\#\#\#\#
\STATE [\texttt{Iid1}, \dots, \texttt{Iidn}] = \texttt{ExposureSeqGen}(\texttt{UidX}, $n$)
\STATE \#\#\#\#\#\#\textbf{PREPARE MARM INPUT}\#\#\#\#\#\#
\STATE $\mathbf{IidY}$ = \texttt{ModelSparseFeatureLookUp}(\texttt{IidY})
\STATE $[\mathbf{Iid1}, \dots, \mathbf{Iidn}]$ = \texttt{ModelSparseFeatureLookUp}([\texttt{Iid1}, \dots, \texttt{Iidn}])
\FOR {$i$ from $1 \rightarrow {L}$}
\STATE $[\texttt{Iid1}^i_{\texttt{UidX}},\dots, \texttt{Iidn}^i_{\texttt{UidX}}]$ = \texttt{UserItemDepthHash}(\texttt{UidX}, [\texttt{Iid1}, \dots], $i$)
\STATE $[\mathbf{Iid1}^i_{\texttt{UidX}},\dots, \mathbf{Iidn}^i_{\texttt{UidX}}]$ = \texttt{MARMCacheLookUp}($[\texttt{Iid1}^i_{\texttt{UidX}},\dots]$)
\ENDFOR
\STATE \#\#\#\#\#\#\textbf{MARM FORWARD}\#\#\#\#\#\#
\STATE $\mathbf{IidY}_{\texttt{UidX}}^1 = \texttt{Target-Attention}(\mathbf{IidY}, [\mathbf{Iid1}, \dots, \mathbf{Iidn}])$\\
\FOR {$i$ from $1 \rightarrow L$}
\STATE $\mathbf{IidY}_{\texttt{UidX}}^{i+1} = \texttt{Target-Attention}^{i}(\mathbf{IidY}_{\texttt{UidX}}^{i}, [\mathbf{Iid1}_{\texttt{UidX}}^{i}, \dots, \mathbf{Iidn}_{\texttt{UidX}}^{i}])$
\ENDFOR
\STATE \#\#\#\#\#\#\textbf{SAVE RESULTS TO CACHE MEMORY}\#\#\#\#\#\
\STATE $[\texttt{IidY}^1_{\texttt{UidX}},\dots, \texttt{IidY}^{L}_{\texttt{UidX}}] = \texttt{UserItemDepthHash}(\texttt{UidX}, \texttt{IidY}, [1,\dots,L])$
\STATE $\texttt{MARMCacheSave}([\texttt{IidY}^1_{\texttt{UidX}},\dots,\texttt{IidY}^{L}_{\texttt{UidX}}], [\mathbf{IidY}_{\texttt{UidX}}^1, \dots, \mathbf{IidY}_{\texttt{UidX}}^{L}])$
\ENDWHILE
\end{algorithmic}
\end{algorithm}

\section{Cache Scaling-Laws}
%
In this section, we will explore a new Scaling-Laws to unlock the future of RecSys, the MARM's Cache Scaling-Laws.

\subsection{Evaluation Setting}
\subsubsection{Datasets}
To evaluate the effectiveness of our MARM, we conduct detailed experiments at Real-world RecSys.
%
%
%
To be specific, this short-video scenario includes approximately 30 million users and 62 million short videos, with each user reading an average of 133 short videos per day.
We collect past 6 months' logs of this scenario, and then conduct the common \textbf{multi-task learning with naive MARM setting (without SIM)} in streaming training manner.

\subsubsection{Metrics}
 For brevity, we present GAUC and training loss of \textbf{long-view prediction task only} to reflect model's performance. 
GAUC, a metric most closely related to online effectiveness, employs a weighted average of the AUC for each user, where the weights are determined by the number of samples for that user:
\begin{equation}
\small
\begin{split}
\texttt{GAUC} = \sum_{u} w_u \texttt{AUC}_u \quad \texttt{where}\ \ w_u = \frac{\texttt{logs}_u}{\texttt{total\_logs}},
\end{split}
\label{gauc}
\end{equation}
\textbf{We show the average GAUC and training loss of the last day}.
%



\subsection{Discussion of Cache Scaling Laws}\label{sec:scaling-law} 
%
Before going on, we first present an important concept, the \textbf{cache size} $\mathcal{C}=L*n*d$, where \textit{L} is the attention layer depth of MARM, \textit{n} is the length of the item sequence, and \textit{d} is the representation dimension.
It is worth noting that both the computational complexity of the cache size and the amount of storage resources are linearly proportional to $\mathcal{C}$. This means that the size of $\mathcal{C}$ is linearly positively correlated with the online latency of the recommendation system's MARM module, as well as with all resources involved in training, inference, and storage, ensuring that the increase in FLOPs is offset by the efficiencies gained from the cache techniques.
%
%





\begin{figure}
    \centering
    \subfigure[Model performance of dimension \textit{d} and length $n$ scaling up while keeping \textit{L}=0]
    {
    \includegraphics[width=8cm,height=3.11cm]{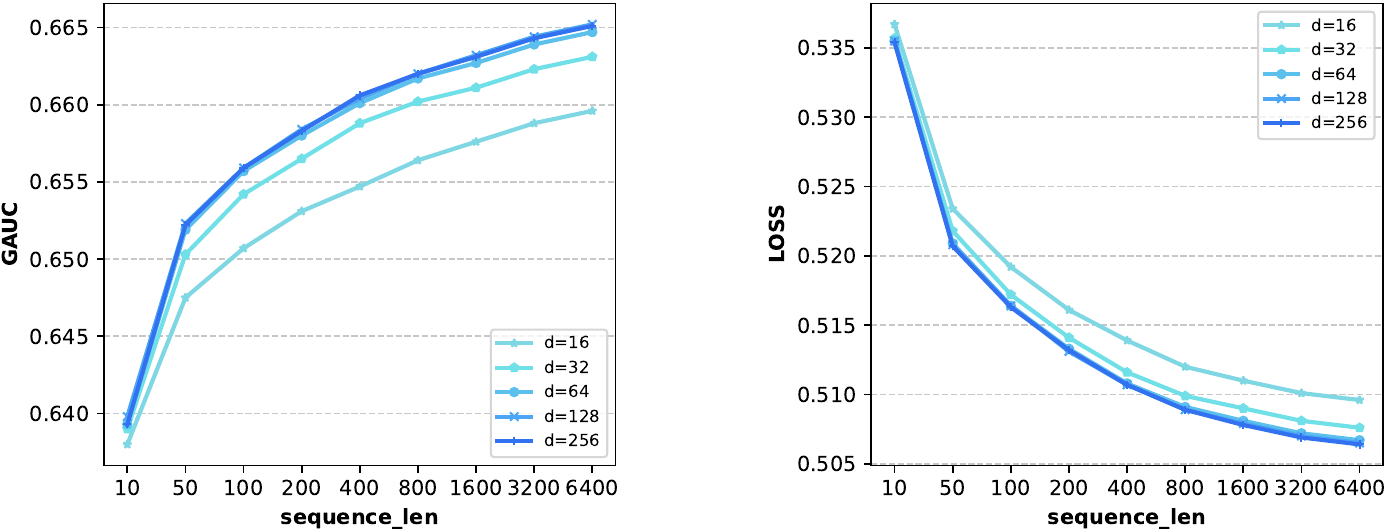}
    \label{subfig1} 
    }
        \subfigure[Model performance of depth \textit{L} and length \textit{n} scaling up while keeping \textit{d}=128]{
    \includegraphics[width=8cm,height=3.11cm]{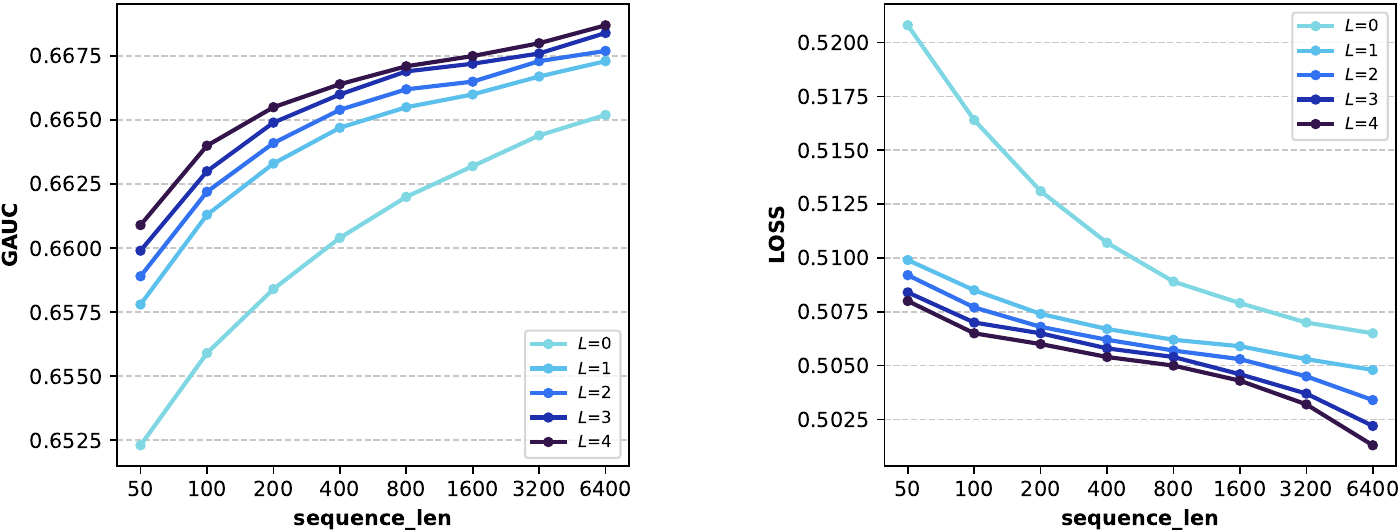}
    \label{subfig2}
}
        \subfigure[Model performance of cache size $\mathcal{C}$ scaling up]{
    \includegraphics[width=8cm,height=3.11cm]{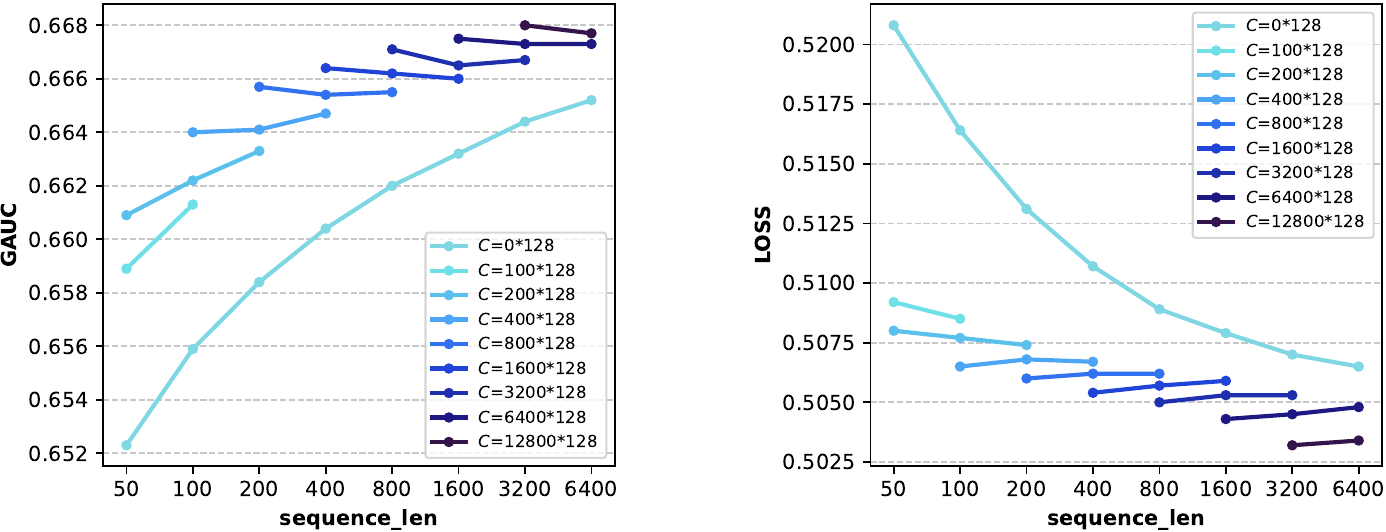}
    \label{subfig3}
 }
    \caption{Model performance of MARM scaling up.}
    \label{fig:enter-label}
    \vspace{-3mm}
\end{figure}

%
%

\subsubsection{Scaling up the dimension $d$ and sequence length $n$}
Generally speaking, the influence of the representation dimension \textit{d} on modeling effectiveness is related to the features used and the complexity of the dataset.
To prevent a combinatorial explosion, we performed a grid search on the representation dimension \textit{d} and sequence length \textit{n} without our MARM module (i.e., $L=0$, equivalent to DIN, with one layer of target attention).
As shown in Figure~\ref{fig:enter-label}(a), we can see that once the $d$ reaches 128, further increases have a minimal impact on performance.
Therefore, in the following, we fix the dimension $d=128$ to conduct more comprehensive analyses.

\subsubsection{Scaling up the depth $L$ and sequence length $n$}
%
As shown in Figure~\ref{fig:enter-label}(b), we can see a significant positive correlation between increasing MARM depth $L$ and sequence length $n$. 
Additionally, it is observed that when the length of the user's behavior sequence increases, even reaching 6400, increasing the MARM depth $L$ still could lead to a notable improvement in model performance, which validates that caching the intermediate results is a potential way to unlock a new direction to build better RecSys.
%
%
%

\subsubsection{Scaling up the cache size $\mathcal{C}$}
Based on the above studies, we further analyzed the relationship between the model's performance with the cache size \textit{C}. 
%
Specifically, we draw several lines to represent \textbf{same Cache size and FLOPs variants but with different settings, e.g., 400*128 has three variants 4*100*128 $\leftrightarrow$ 2*200*128 $\leftrightarrow$ 1*400*128}.
As shown in Figure 5(c), there is a noticeable power-law improvement trend in model performance when the cache size is increased.
%
%
An interesting phenomenon is that when the cache size is small, increasing the sequence length $n$ yields significantly better results than increasing the depth $L$, e.g., 200*128. 
However, \textbf{once the cache size reaches a certain level}, the effects of increasing sequence length and MARM depth become comparable, which leads to an excited observation that \textbf{models with \textbf{same enough Cache size} will show similar performance}, e.g., 6400*128.
%

\subsection{Real-World Experiments}

\begin{table} [t]\tiny
\footnotesize
\centering
\setlength{\tabcolsep}{10pt}
\setlength{\extrarowheight}{2pt}{
\centering
\caption{Offline comparison with SOTAs. Each module is added individually to the baseline model. The arrows in the data on MARM represent a comparison of the FLOPs of the uncache version of MARM. The best and runner-up results are highlighted in bold/underlined.}
\vspace{-10pt}
\resizebox{8cm}{!}{ 
    \begin{tabular}{lccc} 
         \toprule
         &  AUC ($mean \pm std $ )& GAUC ($mean \pm std $) &FLOPs\\
         \midrule
 Baseline& $0.8233 \pm 0.0001$&$0.7093 \pm 0.00005$ &\\
 \midrule
 DIN& $0.8235 \pm 0.00019$&$ 0.7010 \pm 0.00012 $ &3.3M\\
 SIM Soft& $ 0.8241 \pm 0.00004 $&$ 0.7109 \pm 0.00007 $ &4.6M\\
 TWIN& $ 0.8249 \pm 0.00005$&$ 0.7142 \pm 0.00009 $ &90M\\
 \underline{TWIN V2}& \underline{$0.8251 \pm 0.00023$}&\underline{$ 0.7147 \pm 0.00007 $} &150M\\ 
         \midrule
         MARM(L=1)&  $0.8256 \pm 0.00008 $& $0.7157 \pm 0.00006 $ &127M $\leftarrow$ 2.7B*\\ 
         MARM(L=2)&  $0.8262 \pm 0.0004$& $0.7169 \pm 0.00012 $ &164M $\leftarrow$ 5.4B*\\ 
         \textbf{MARM(L=4)}&  $\mathbf{0.8270 \pm 0.0003}$& $\mathbf{0.7183 \pm 0.00008}$ &238M $\leftarrow$ 10.5B*\\
\midrule
         Improvement& $0.19\%$& $0.36\%$ &\\
         \bottomrule
    \end{tabular}
    
}
\label{individual}
}
The FLOPs marked with * indicates the un-cache version MARM needed.
\vspace{-3mm}
\end{table}


\subsubsection{Compared Approaches Details}
%
%
%
%
%
Generally speaking, MARM can be seen as a user sequence modeling module; therefore, we select the following strong methods to verify MARM's ability:
%
%
\begin{itemize}[leftmargin=*,align=left]
    \item \textbf{Baseline}: A model using multi-task MoE structure with features of users, videos, and statistics. User sequence information is the result of sum pooling of user's short-term historical behavior.
    \item \textbf{DIN}\cite{zhou2018deep}: The most commonly used algorithm for modeling user short-term historical behavior, utilizing a target attention mechanism. The user history length we are using here is 50.
    \item \textbf{SIM Soft}\cite{pi2020search}: A two-stage modeling approach with GSU-ESU, where GSU uses pre-trained multi-modal embeddings of videos to calculate inner product and select top-k most relevant videos from user's history. The user history length used here is 15000.
    \item \textbf{TWIN}\cite{chang2023twin}: A two-stage modeling approach under SIM architecture, aligning computation methods of GSU and ESU to enhance consistency between them. User history length used is 15,000.
    \item \textbf{TWIN V2}\cite{si2024twin}: This approach uses hierarchical clustering to reduce scale of ultra-long user historical lengths, followed by modeling using TWIN method. User history length used is 100,000.
    \item \textbf{HSTU*}\cite{zhai2024actions}: This method employs an HSTU-style multi-layer self-masked attention mechanism to model user history (time complexity $\mathcal{O}(n^2*d)$). Due to computational limitations, each historical item only uses a few features: item ID, author ID, tag, and user feedback, with 2000 history length and 4 depth. \textbf{Its FLOPs can be seen as the un-cache version of MARM}. 
    \item \textbf{MARM}: Using the MARM with SIM approach, \textbf{the first layer reuses the existing TWIN module}, we then stack MARM modules, with each MARM module storing a length of 6000 and a maximum attention depth \textit{L} of 4\footnote{Actually, the MARM framework can be stacked based on any existing user sequence modeling module, and of course, we acknowledge that the foundational module will significantly impact the final results.}.
\end{itemize}
Based on the above approaches, we conducted two types of experiments with respect to the conventions of academics and industry: 
\begin{itemize}[leftmargin=*,align=left]
\item \textbf{Individual experiment}: This way only contains the baseline model and the corresponding method modification.
\item \textbf{Ensemble experiment}: After a modification is verified to be effective, such method will be fused with later modifications.
\end{itemize}

\subsubsection{Individual Experiment Performance Comparisons}
In this section, we introduce different sequence modeling modules into the baseline model, and the experimental results are shown in Table~\ref{individual}.
MARM achieves the best performance compared to all the baselines. It is worth noting that although MARM does not use a user history length in the range of 100,000 like TWINV2, but only 6000, its performance still significantly outperforms TWINV2. 
We also list the FLOPs metrics for each module. It should be noted that in the two-stage modeling architecture of SIM, which consists of GSU and ESU, although the sequence length of the ESU during training is only a few hundred, for fairness, we include the computational cost of the previously calculated GSU part when calculating FLOPs.
Thanks to MARM's ability to scale linearly in depth, we can expand the attention depth, not just the sequence length of the users. Furthermore, we can observe that with the increase in attention depth, MARM still shows significant improvement.


\begin{table} [t]\tiny
\footnotesize
\centering
\setlength{\tabcolsep}{10pt}
\setlength{\extrarowheight}{3pt}{
\centering
\caption{Ensemble comparison with SOTAs. If each module brings a significant gain in confidence, it will be added one by one. Taking TWIN V2 as an example, it includes the Baseline, DIN, SIM Soft, TWIN, and TWIN V2.}
\vspace{-10pt}
\resizebox{8cm}{!}{ 
    \begin{tabular}{lccc} 
         \toprule
         &  AUC ($mean \pm std $ )& GAUC ($mean \pm std $) &FLOPs\\
         \midrule
 Baseline& $0.8233 \pm 0.0001$&$0.7093 \pm 0.00005$ &\\
 \midrule
 DIN& $0.8235 \pm 0.00019$&$ 0.7010 \pm 0.00012 $ &3.3M\\
 $\quad$+SIM Soft& $ 0.8251 \pm 0.00004 $&$ 0.7111 \pm 0.00008 $ &7.9M\\
 $\quad\quad$+TWIN& $ 0.8260 \pm 0.00005$&$ 0.7156 \pm 0.00006 $ &94.6M\\
 $\quad\quad\quad$+TWIN V2& $0.8268 \pm 0.00023$&$ 0.7167 \pm 0.00011 $ &154.6M\\
\midrule
$\quad\quad\quad\quad$+HSTU*& $ 0.8268 \pm 0.00017 $&$ 0.7168 \pm 0.00021 $ &1.04B\\
\midrule
         $\quad\quad\quad\quad$+MARM(L=1)&  $0.8270 \pm 0.00023 $& $0.7180 \pm 0.00020 $ &191.6M$\leftarrow$ 2.77B*\\ 
         $\quad\quad\quad\quad$+MARM(L=2)&  $0.8272 \pm 0.00019$& $0.7191 \pm 0.00021 $ &228.6M$\leftarrow$ 5.47B*\\ 
         $\quad\quad\quad\quad$+\textbf{MARM(L=4)}&  $\mathbf{0.8286 \pm 0.00013}$& $\mathbf{0.7210 \pm 0.00011}$ &302.6M$\leftarrow$ 10.57B*\\
\midrule
         Improvement& $0.22\%$& $0.43\%$ &\\
         \bottomrule
    \end{tabular}
}
\label{ensemble}
}
\vspace{-3mm}
\end{table}

\begin{figure}[t]
    \centering
    \includegraphics[width=8cm,height=5cm]{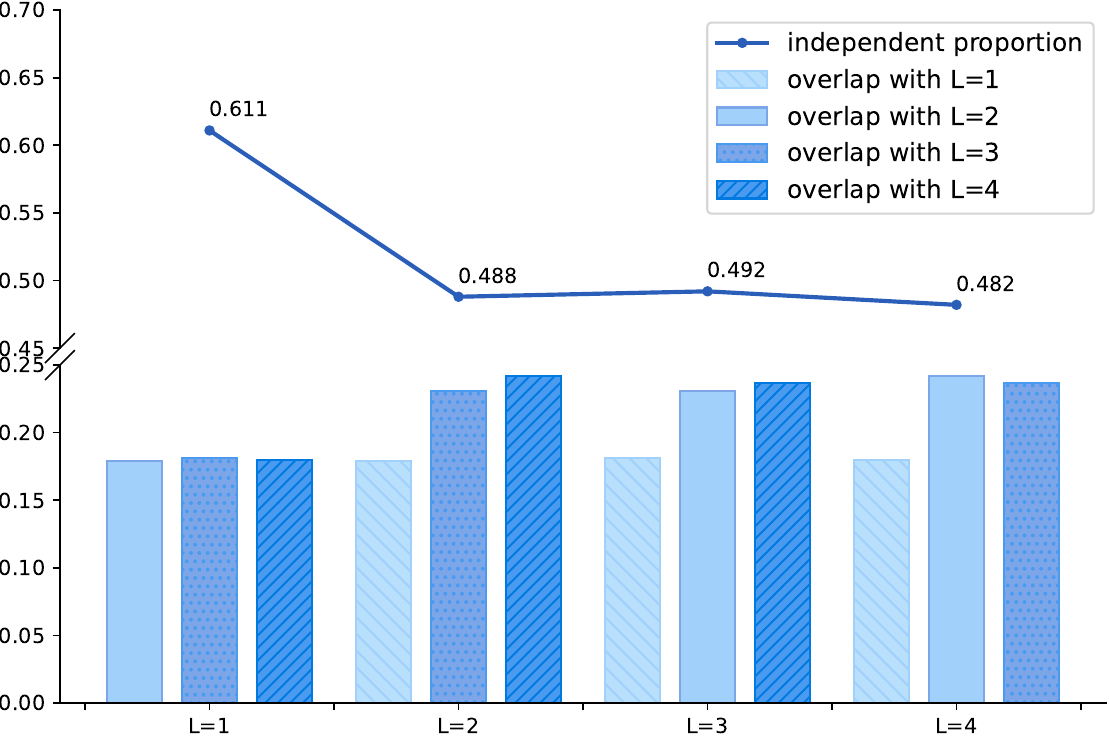}
    \caption{GSU Top-k overlapped rate between different layers.}
    \label{fig:simoverlap}
    \vspace{-3mm}
\end{figure}
\begin{table*}[t]
\centering
\caption{Online A/B testing results of Short-Video services}
\setlength{\tabcolsep}{2pt}{
\begin{tabular}{c|llccccccc}
\toprule
\multirow{2}{*}{Stages}                     &   \multicolumn{5}{c}{Retention\&Watch Time Metrics}& \multicolumn{4}{c}{Interaction Metrics}\\
\cmidrule(r){2-6}  \cmidrule(r){7-10}&  1D Retention&3D Retention&Average Watch Time & Watch Time   & Video View                                      & Like                   & Comment                   & Forward    &Follow                 \\
\midrule
Retrieval  &   +0.02\%&+0.009\%&+0.489\% &	+0.456\% &	+0.463\% &	+0.167\% &	+0.675\% &	-0.550\% &	+0.288\% \\ 
Cascading &   +0.01\%&+0.01\%&+0.276\% &	+0.260\% &	+0.086\% &	+0.100\% &	-0.017\% &	+0.495\% &	+0.870\% \\
Ranking &   +0.07\%&+0.13\%&+1.314\% &	+1.370\% &	+0.103\% &	+0.605\% &	-0.669\%&	-0.114\% &	-0.227\%\\
\bottomrule
\end{tabular}}
\label{mainonline}
\end{table*}

\subsubsection{Ensemble Experiment Offline Performance}
This section aims to answer the following question: could MARM module further improve a model that already incorporates various long-term and short-term user behavior modeling?
Therefore, we conducted another set of ensemble experiments and the results are shown in Table~\ref{ensemble}.
Specifically, we sequentially added DIN, SIM Soft, TWIN, TWIN V2, and MARM modules to the baseline, observing improvements brought by them. 
From Table~\ref{ensemble}, after adding a series of long-term and short-term modeling modules, adding a simplified HSTU-style module provided no significant improvements, possibly because the scale of the features we used, attention depth, or historical length did not reach a critical mass. 
%
%
Besides, another reason might be that our MARM could save more interaction sample knowledge. While a pure item sequence has only a few tens of attributes, MARM's cached results are always thousand-dimensional compressed target item query information, which achieves potential sample communication within MARM.
In addition, directly adding an HSTU-style module significantly increased the computational burden on our model. 
In our scenario, when we add MARM to a system that includes several long-term and short-term sequence modeling models, there is still a significant improvement in accuracy. 
\textbf{MARM is a very practical module that can be added to most recommendation models at a very controllable cost.}

\subsubsection{GSU searched Top-K Overlapped Rate of MARM}
First, we aim to understand how the attention depth of MARM leads to sustained improvements.
In the framework of MARM with SIM, each layer of MARM Block models GSU and ESU phases like TWIN. 
The GSU returns top-k user histories most relevant to target item. 
As shown in Figure~\ref{fig:simoverlap}, we analyzed the overlap of the histories returned by GSU of four-layer MARM Block and found that the proportion of independence for each layer is relatively high, exceeding 50\%. 
Moreover, the direct overlap between any two layers is \textbf{less than 20\%}. This indicates that each layer of MARM focuses on different historical content, forming a high-level expression of interests.
In addition, we elaborate on the innovation and effectiveness analysis of the cache strategy used in MARM in the Sec ~\ref{sec:opt-tech}.

\subsubsection{MARM Cost Discussion}
%
As an equivalent form of an un-cached MARM, HSTU follows a first sample-aggregation then learning paradigm to alleviate calculation pressure, its overall computational complexity for a user with a lifelong historical behavior is about $O(L*n^2*d*(\frac{N}{r}))$, where \textit{N} denotes total interaction amount of a user, and $\textit{r}<<n$ is aggregation level of user samples, meaning that every \textit{r} user samples are aggregated together for training. 
In practice, a larger aggregation \textit{r} typically reduces the modeling complexity but can lead to delays in user feedback entering the model training, and easily results in a decline in online metrics, creating a trade-off between cost and effectiveness.

%
Compared to above, HSTU applies a high-FLOPs masked-self-attention; our MARM is a more low-resource solution, needing no aggregation of training samples for user modeling.
For overall learning process of a user, the computational complexity of MARM is about $O(L*n*d)$.
As a replacement, MARM does require additional storage, which we previously referred to as cache size $\mathcal{C}$. 
In our scenario, with attention depth $L=4$ and sequence length $n=6000$, \textbf{the storage used by MARM is 60TB}.
\textbf{The combined storage and increased computational overhead is approximately $1/8$ that of a direct multi-layer self-attention approach. }

For the deployed MARM version with parameters $L=4, n=6000,d=128$, serving 30M DAU, MARM uses only 100 A10 GPUs ( ¥3.5M/year ) + 60TB Storage ( ¥1.2M/year ); these resources encompass all aspects of training and inference. The sota HSTU would require at least 10 times computing resources of MARM, i.e., 1000 A10 GPUs( ¥35M/year ). 
The storage cost is roughly 1/3 of the additional computation cost, which is why MARM's approach of exchanging storage for computation is so cost-effective.
\begin{figure}[t]
  \centering
  \includegraphics[width=8cm,height=4.5cm]{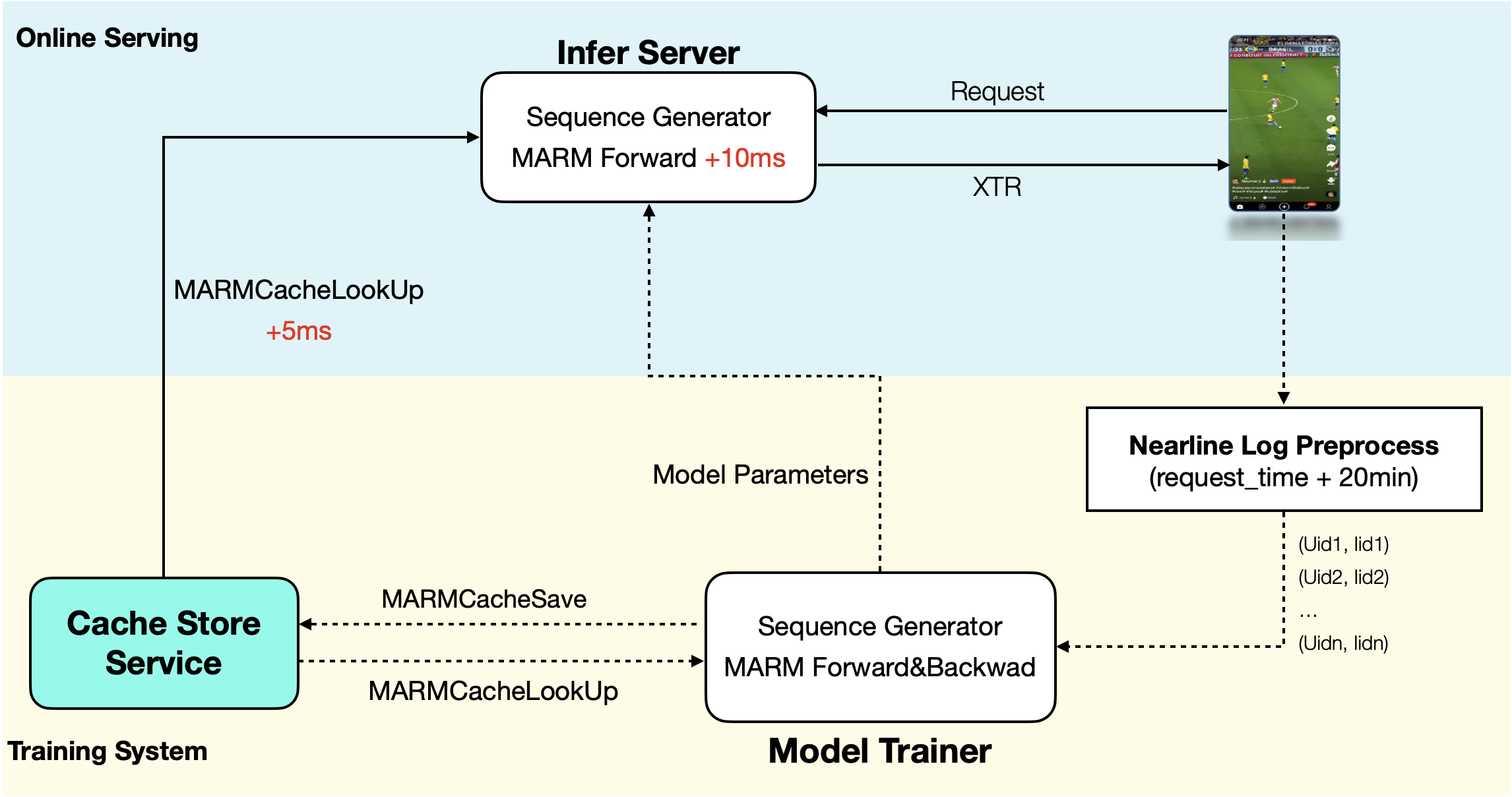}

  \caption{The online ranking time cost of the deployed version of MARM is approximately 15ms, with the embedding pulling taking about 5ms and the forward computation for inference taking about 10ms. }
  \label{time_cost}
\end{figure}
In addition, we show in Figure~\ref{time_cost} additional time cost that MARM brings to inference service. Compared to performance improvement provided by MARM, the time cost increase (+15ms) is a worthwhile trade-off: on par with TWIN (+15ms) and much better than HSTU (+40ms).

\subsubsection{Online Results}
To quantify the contribution of the MARM model to real-world RecSys, we implemented it across the Retrieval, Cascading, and Ranking stages, validating its effectiveness through our online A/B testing system.
The model's performance was assessed based on key metrics, including core playtime metrics and interaction metrics such as average watch time per use and the number of likes.
Table~\ref{mainonline} presents the online results of MARM across various recommendation stages.

Our app's retention metrics are positive, indicating that the launch of MARM has resulted in a better user experience. Notably, during multi-stage, MARM achieved a significant increase in average user app usage time of +2.079\%, highlighting its crucial role in enhancing user watch time. While there were slight negative impacts on Comment/Forward/Follow, significant gains were achieved in user interactions through Like. Since the primary goal of the recommendation system is to maximize average watch time, the online results are still within a reasonable range of substitution. MARM has already achieved significant business benefits in real-world recommendation systems and has been deployed to serve tens of millions of users for over a year. Moreover, through continuous iterations, MARM has brought over a 5\% increase in user app usage time to our business.

\section{Cache Strategy Discussion}\label{sec:opt-tech}
\subsubsection{Innovation Description}
%
Actually, the NLP KV cache technique is typically applied in the \textbf{inference stage} of autoregressive LLM to reduce the time complexity of generating the next token. 
Nevertheless, our MARM cache technique is used \textbf{in both the training stage and the inference stage}.
Overall, we have the following insightful and crucial innovations:
\begin{itemize}[leftmargin=*,align=left]
   \item{\textbf{Cache Generation}}: Different from NLP KV cache technique which generates cache results while predicting the next token in model inference process, our MARM generates the cache results in training process, which aims to save all user-item behavior pattern information.
    In our MARM inference, which is shared with training-produced cache results, we focus on utilizing the saved cache results for better prediction.

    \item{\textbf{Cache Life-Cycle}}: 
    In particular, the NLP KV Cache for current sentence generation only needs temporary life-cycle storage in GPU memory.
    Consequently, once the sentence is generated completely, its corresponding cached data is no longer necessary, and the cache results will be deleted.
    Unlike language sentences that have endpoints, in streaming recommendation systems, user-item behaviors are generated sequentially, and our users do not have an `endpoint' to finish.
    Thus, our MARM Cache should have a long-term life-cycle to store user interests, to ensure that our model can access those cached results at any time to provide high-quality recommendation accuracy.
    %
   

    \item{\textbf{Cache Scope}}: 
    In NLP KV Cache, it focuses on caching all tokens' transformed Key-results and Value-results, waiting for the incoming Query to aggregate them.
    In our MARM, we do not cache the transformed Key-results and Value-results, but cache the final calculated Query outputs, which could reduce the overall storage requirement, conserving resources efficiently.
    \item{\textbf{Complexity and Resoruces}}:
    In the fields of NLP and recommendation, similar generative models like HSTU exhibit a computational complexity of  $\mathcal{O}(n^2)$
    for both training and inference. Although these models utilize techniques such as kv cache during inference, they still need to compute the self-attention for the entire sequence at least once. In contrast, MARM achieves a computational complexity of $\mathcal{O}(n)$ for both training and inference because it caches the computation results for each layer in the sequence. However, MARM does incur the additional burden of requiring an external storage resource.
\end{itemize}

\subsubsection{Effectiveness Analysis} \label{effective_analysis}
Compared to computation methods using complete transformer structures like HSTU, the cache-based method of MARM is essentially equivalent in terms of computation, with the main differences being the following two points:
\begin{itemize}[leftmargin=*,align=left]
    \item Since the key/value pairs of each layer in MARM are read from the cache, this means that the key/value pairs between two consecutive attention layers in MARM are not connected as nodes in the computation graph, only the query is connected. This implies that during backpropagation, the gradient does not pass between the key/value pairs; it only passes from the query part of the later layer to the key/value part of the earlier layer.
    \item The cache in MARM is updated only once, specifically during the computation for a particular sample when it is trained and written to the cache. The portions written to the cache will be frozen, while certain model parameters retained in the computation graph, such as the Q, K and V mapping matrices and the FFN parameters for each layer, will continue to be updated.
\end{itemize}
We would like to explain why the two differences mentioned above do not significantly impair the performance of the MARM architecture. 
In Streaming RecSys, items that have already been exposed to a specific user will not be exposed again. Therefore, recommendation scenarios typically train using one epoch of streaming data, and repeatedly training on the same sample usually does not lead to performance improvement and may even harm the recommendation effectiveness. As a result, recommendation models remain in a state of high generalization after convergence, allowing them to handle newly emerging items and accurately model item characteristics.

Regarding the first difference, if the model already exhibits strong generalization and can accurately model the user-item pairs, it can be stored and used directly without the need for extensive adjustments during subsequent training. Instead, it is sufficient to adjust only certain parameters retained in the computation graph.

For the second difference, our analysis of the lower-level model parameters in recommendation models reveals that their changes are relatively slow, whereas the parameters closer to the logits change more rapidly. Therefore, freezing these parameters will not quickly lead to drift. Moreover, in the streaming recommendation scenario, 
each item often has its own lifecycle, and using the computed results from frozen parameters during its lifecycle may not be significantly worse than using the latest model parameters. This is why the cache-based MARM method still shows excellent performance and scalability in streaming recommendation scenarios.

\section{Conclusion}
We propose the MARM, a multi-layer recommendation model that utilizes Memory Cache to accelerate inference. In recommendation scenarios, computational complexity is a significant performance constraint. We use caching to store partial computation results of complex models, reducing the complexity of single modeling from $\mathcal{O}(n^2*d) \rightarrow \mathcal{O}(n*d)$. Based on the MARM framework, we can extend sequence modeling from single-layer target attention to multi-layer with linear resource consumption, significantly breaking through the computational bottleneck of recommendation models and enabling modeling of users' life-long histories. We explored the scaling laws within the MARM framework, confirming a proportional relationship between cache size and recommendation performance. Furthermore, our MARM method can seamlessly integrate with existing recommendation models, including fine ranking, coarse ranking, and retrieval. 

\section{GenAI Usage Disclosure}
In this paper, we only utilize the AI tools to fix our grammar mistakes.
The research motivation, methodology, and experimental results are entirely derived from first-hand experiments and analysis conducted in real-world business scenarios. 
All data and observations have been rigorously validated by online A/B tests and offline analyses.

\balance
\nocite{*}
\bibliographystyle{ACM-Reference-Format}
\bibliography{marm.bib}

\end{document}